\documentclass[fleqn,10pt]{wlscirep}
\usepackage[utf8]{inputenc}
\usepackage[T1]{fontenc}

\usepackage{color}
\usepackage{graphicx}

\newcommand{\OD}[2]{\frac{\textrm d #1}{\textrm d #2}}

\title{
Manipulating Collective Opinion through \\ Social Network Intervention
}

\author[1]{Shigefumi Hata}
\author[2]{Renaud Lambiotte}
\author[3]{Hiroya Nakao}
\author[4,5*]{Ryota Kobayashi}
\affil[1]{Graduate School of Science and Engineering, Kagoshima University, Kagoshima, 890-0065, Japan}
\affil[2]{Mathematical Institute, University of Oxford, Woodstock Road, Oxford, United Kingdom}
\affil[3]{Department of Systems and Control Engineering and Research Center for Autonomous Systems Materiaology, Institute of Science Tokyo, Meguro, 152-8552, Japan}
\affil[4]{Department of Complexity Science and Engineering, The University of Tokyo, Kashiwa, 277-0882, Japan}
\affil[5]{Mathematics and Informatics Center, The University of Tokyo, Bunkyo, 113-8656, Japan}

\affil[*]{r-koba@edu.k.u-tokyo.ac.jp}

\begin{abstract}
Social media platforms have transformed the dynamics of collective opinion formation, enabling rapid, large-scale interactions while simultaneously exposing online discourse to polarization and manipulation. Traditional models of opinion dynamics often predict convergence to a consensus, yet empirical evidence consistently highlights persistent polarization and radicalization, especially on contentious issues. This paper analytically investigates a mathematical model that captures the complex interplay of polarization, radicalization, and consensus within networked societies. 
By analyzing the emergence and stability of opinion clusters, we identify critical thresholds marking phase transitions in collective behavior, interpreted via a stability landscape. 
We further explore network-based interventions to manipulate the collective opinion, revealing that reducing inter-agent interactions can lead to unintended, irreversible shifts in opinion distributions. 
Our results underscore the dual-edged nature of intervention strategies, offering theoretical insight into the fragility and manipulability of public opinion in digital environments.
\end{abstract}  
\begin{document}

\flushbottom
\maketitle
%
%
\thispagestyle{empty}

\section*{Introduction}  

Political polarization, often described as ideological polarization, occurs when people’s views on political issues split into opposing camps. Over the past decade, researchers from many fields have explored this growing divide using methods that range from large-scale data analysis~\cite{conover2011political,mccright2011politicization,boxell2017greater,casal2021polarization,garimella2017long,bottcher2020great,flamino2023political,lee2018does} and field experiments~\cite{bail2018exposure,allcott2020welfare,mosquera2020economic} to computer simulations and agent-based models~\cite{deffuant2000mixing,axelrod2021preventing,lu2025agents,sasahara2021social,baumann2020modeling,santos2021link,perra2019modelling}. In the United States, polarization has systematically intensified around issues like climate change, gun control, and healthcare~\cite{mccright2011politicization,boxell2017greater}, while recent evidence shows that opinion radicalization is also rising across Europe~\cite{casal2021polarization}. As these divides deepen, opportunities for compromise shrink, increasing the risk of political deadlock and threatening the well-functioning of democratic institutions~\cite{Levitsky2018}.

Researchers have identified multiple drivers of political polarization, including the rise of digital and social media, widening income inequality, elite polarization, and demographic shifts~\cite{fiorina2008political,prior2013media,kubin2021role,van2021social,arora2022polarization,Azzimonti2023}.
Although the evidence remains inconclusive, social media is widely regarded as a significant contributor to polarization dynamics~\cite{kubin2021role,van2021social,arora2022polarization,thurner2025more}.
Panel data analyses~\cite{lee2018does} and large-scale field experiments~\cite{bail2018exposure,allcott2020welfare,mosquera2020economic} provide indirect support for the hypothesis that online platforms amplify political divisions. 
Earlier research has largely explained polarization through mechanisms such as homophily, that is the tendency for individuals to preferentially associate with those who are similar~\cite{mcpherson2001birds,vazquez2008generic}, and the formation of communities within social networks~\cite{lambiotte2007majority}. 
Mathematical modeling further shows that when agents with similar views mutually reinforce one another, combined with homophily, the resulting dynamics can reproduce empirically observed phenomena such as radicalization, polarization, and the emergence of echo chambers~\cite{baumann2020modeling,santos2021link}.  
Moreover, mathematical models have been employed to propose strategies for mitigating political polarization~\cite{axelrod2021preventing,lu2025agents}. 
Since conducting large-scale empirical experiments is often impractical, research based on mathematical models plays a crucial role. 
While prior studies have primarily focused on user characteristics and the influence of bots, the potential for social network providers to intervene directly in the network structure remains underexplored. In particular, the effects of such structural interventions on collective opinion dynamics have yet to be systematically investigated.

This paper investigates the impact of social network interventions, that is, temporary modifications to the structure of a social network, on political polarization (hereafter, polarization). Specifically, we explore an opinion dynamics model of users within a network and pose two central questions: 1. Can large-scale shifts in users’ collective opinions (e.g., toward radicalization or polarization) be induced through network interventions? 2. If so, what type of user-level information is necessary to enable such manipulation?
To address these questions, we develop a simple mathematical model of opinion dynamics that captures the formation and evolution of opinions within social networks.
Conventional opinion dynamics models generally assume that interactions between agents (i.e., users) promote opinion alignment, leading to eventual convergence on a consensus~\cite{degroot1974reaching,redner2019reality}. However, empirical evidence often shows the persistence of polarization and, in some cases, the emergence of radicalized views on divisive issues.
Departing from the consensus-oriented assumption, our model instead posits that agents with similar opinions reinforce one another’s views~\cite{baumann2020modeling}. This mechanism enables the emergence of more complex opinion states, such as polarization and radicalization, closely reflecting patterns observed in real-world social networks.

Beyond endogenous mechanisms of polarization, increasing attention has turned to the exogenous manipulation of opinion dynamics. A growing body of empirical and theoretical research demonstrates that social media platforms, once celebrated for enhancing democratic discourse, can also be weaponized for large-scale opinion manipulation and social control~\cite{badawy2018analyzing,stewart2019information,mihaylov2018dark,truong2024quantifying}.
Coordinated campaigns employing bots, trolls, and zealots have been shown to distort public opinion, disseminate misinformation, and influence electoral outcomes, often by strategically targeting vulnerable users or communities~\cite{wilder2017controlling}.
Similar techniques have also been applied toward very different goals, such as promoting positive behavior change~\cite{valente2010social}, for instance in substance abuse prevention programs~\cite{valente2007peer}.
From a modeling standpoint, most studies have examined strategies based on the control of agents, capturing the role of bots in shaping collective dynamics, a problem closely related to social influence in online marketing~\cite{romero2020zealotry,yildiz2013binary}. However, to assess the risks of opinion manipulation by social networking service (SNS) providers, it is equally important to evaluate the impact of network-level interventions.
One exception is the study by Chiyomaru and Takemoto~\cite{chiyomaru2022adversarial}, which considers adversarial interventions. Yet, their analysis was restricted to consensus states and did not address polarization dynamics.

Our analysis offers three key contributions.
First, the model is sufficiently simple to allow for an analytical investigation of the conditions under which polarization, radicalization, or consensus emerge. This framework also enables the estimation of critical thresholds, tipping points at which collective opinion dynamics undergo abrupt transitions, providing insight into the mechanisms that drive societies toward polarized or radicalized states.
Second, numerical simulations show that reducing interactions among agents, an intervention that social media platforms could, in principle, implement at scale, can unintentionally alter opinion dynamics in complex and unpredictable ways.
Finally, we find that such interventions may produce irreversible collective shifts in opinion distributions, paradoxically amplifying radicalization rather than mitigating it, and fostering fragmented environments where extreme views proliferate.

\section*{Results}
\begin{figure}[t]
    \begin{center}
        \includegraphics[scale=0.6]{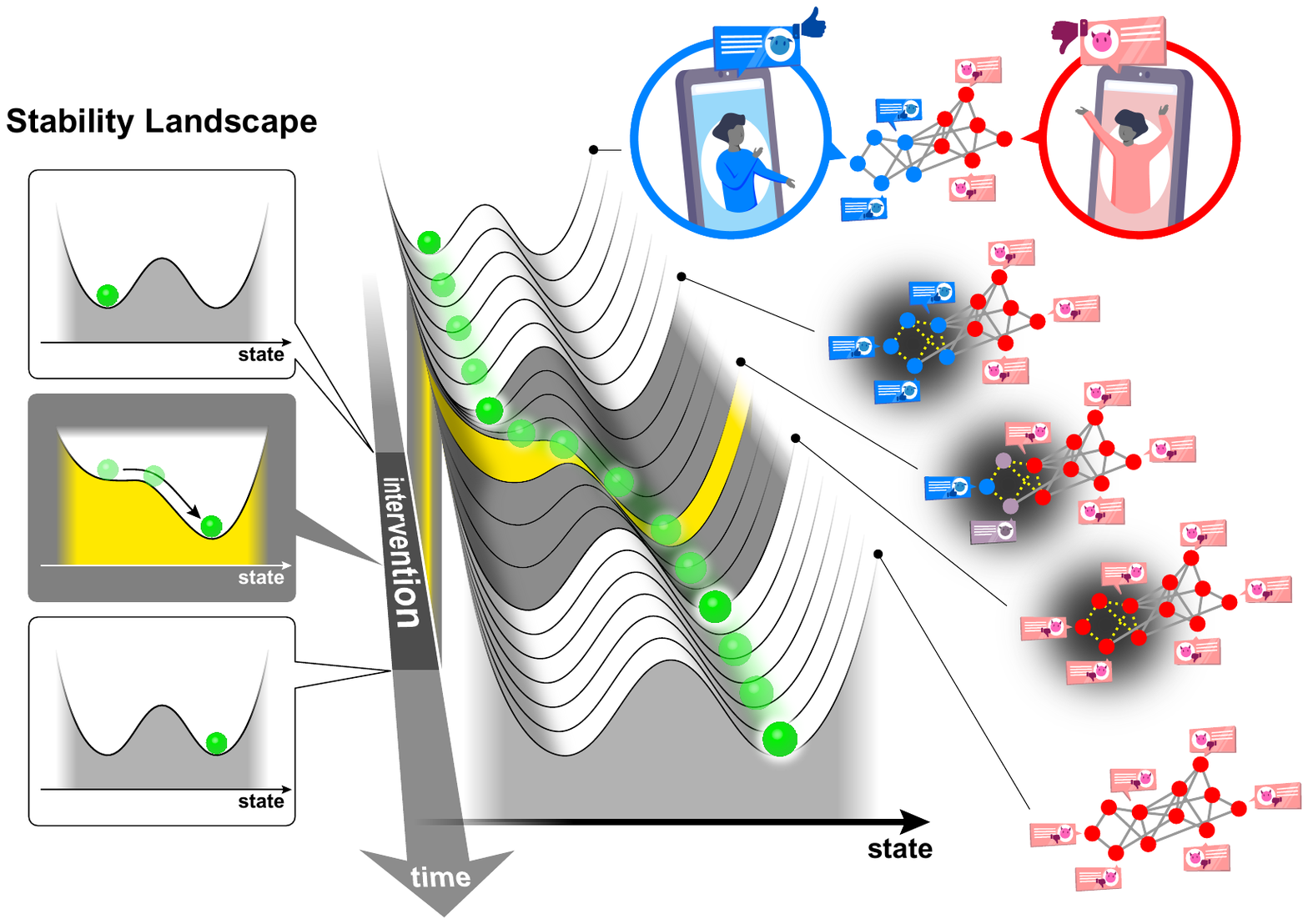}
    \caption{Schematic illustration of this study. \\     
    This study investigates a scenario in which users on social media discuss a particular issue (right). The dynamics of their opinions, i.e., the evolution of all users' views, can be interpreted using a stability landscape: the collective state of opinions is represented as a ball within a potential (center). This study explores how the intervention of the social network may influence the formation of collective opinions (left and right). 
    } 
    \label{fig:Fig_Schematic}
    \end{center}
\end{figure}

This study examines an opinion dynamics model capable of reproducing collective states such as polarization and radicalization. We consider the temporal evolution of users’ opinions through a stability landscape, which has been applied to understand the dynamics of cell biology, ecosystems, and climate change\cite{scheffer2001catastrophic,waddington2014strategy,steffen2018trajectories} (Fig.~\ref{fig:Fig_Schematic}). 
In this framework, users’ opinions correspond to positions on the landscape, while the overall system behavior can be visualized as a ball rolling downhill into a valley, where each valley represents a stable opinion state.
We hypothesize that social network interventions reshape the stability landscape, thereby inducing irreversible shifts in opinion states (Fig.~\ref{fig:Fig_Schematic}). To explore this idea, we proceed as follows: first, we introduce the opinion dynamics model; second, we analyze the phase diagram of opinion states generated by the model; and finally, we interpret the theoretical results in terms of landscape transformations, showing how targeted interventions can manipulate collective opinion states.

\subsection*{Opinion Dynamics Model}
\begin{figure}[t]
    \centering
    \includegraphics[width=0.7 \hsize]{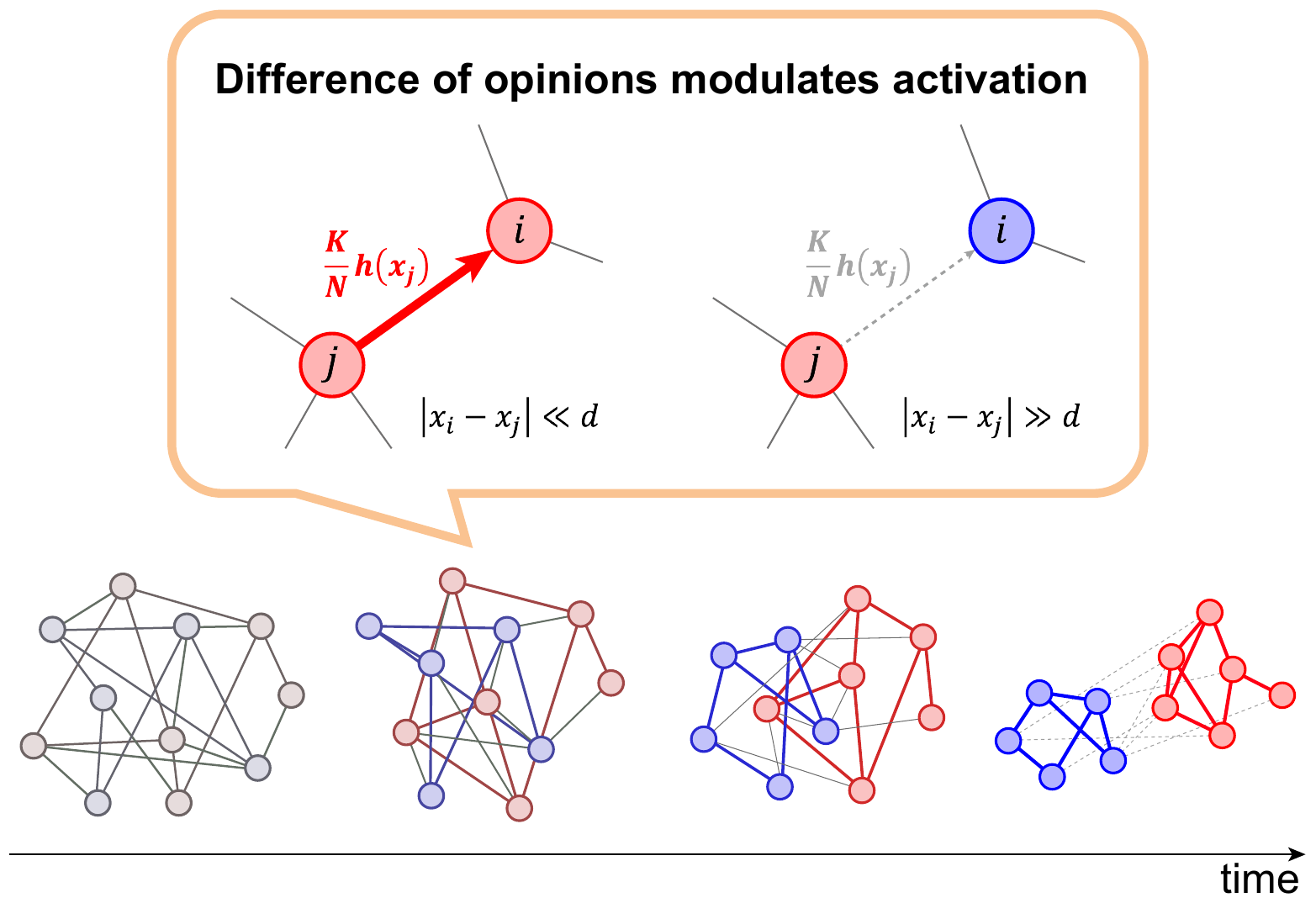}
    \caption{Schematic illustration of the opinion dynamics model (Eq.~\ref{eq_OD}). The color of each node represents the value of the opinion variable, $x_i$, for each user: red, blue, and gray represent that a user is agree, disagree, and neutral, respectively. The influence of the $j$-th user on the opinion of $i$-th user depends on the opinion difference between the users, $|x_i - x_j|$, which is represented by the link width.
    The opinions $\{ x_i \}_{i=1, 2, ..., N}$ thus evolve with the effective coupling strengths between the users. }
    \label{fig:Model}
\end{figure}

We consider a simple mathematical model for opinion formation within a social network. 
This model consists of $N$ individuals (users) who exchange opinions on a topic such as gun control or abortion. 
Each individual $i$ is characterized by its opinion $x_i \in (-\infty, \infty)$, which is represented by a sign indicating the user's stance, i.e., agreement or disagreement, on the issue. The magnitude of $x_i$ represents the strength of the opinion, with a large $|x_i|$ indicating a strong agreement or disagreement with the issue.   
We  assume that individuals have no inherent preference for a particular opinion and that the dynamics are driven solely by social interactions, that is we do not consider external effects such as news. 
The time evolution of the opinions is governed by the system of differential equations 
\begin{equation}
	\frac{\textrm d x_i}{\textrm d t} = -x_i +\frac{K}{N} \sum_{j=1}^N A_{ij} g(|x_i - x_j|; d) h(x_j; \alpha), 
	\label{eq_OD}
\end{equation}
where $K$ is the strength of social interaction between users, proportional to their posting activity, and $A_{ij}$ are the entries of the adjacency matrix of the social network. 
The effect of social interactions on individual opinions is described by $g(|x_i - x_j|; d) h(x_j; \alpha))$ (Fig.~\ref{fig:Model}). 
The homophily function $g(|x_i - x_j|; d)$ describes the tendency of individuals to interact with people with similar opinions, where $|x_i - x_j|$ quantifies the opinion difference between user $i$ and $j$, and $d$  quantifies the opinion tolerance of a user: the user $j$ whose opinion disagreement is greater than $d$ does not influence the opinion of the user $i$ (see Methods for mathematical details). 
The controversy function $h(x_j; \alpha)$ describes the social reinforcement on the issue, where represents the level of controversy surrounding the issue: A weak opinion can have a significant impact on others, especially regarding highly controversial issues (see Methods for mathematical details). 
The proposed model is a variation of the radicalization dynamics proposed in Baumann et al. 2020~\cite{baumann2020modeling}, where homophily is introduced by limiting the influence coming from diverging opinions through the effective coupling $g(|x_i - x_j|)$ instead of imposing a rewiring between diverging opinions. 
Additionally, our model assumes a static network for social interactions rather than the temporal network.
As we will see, these modifications allow us to systematically characterize the asymptotic dynamics of the system.

\subsection*{Phase diagram}
\begin{figure*}[t]
    \centering
    \includegraphics[width= 0.7 \hsize]{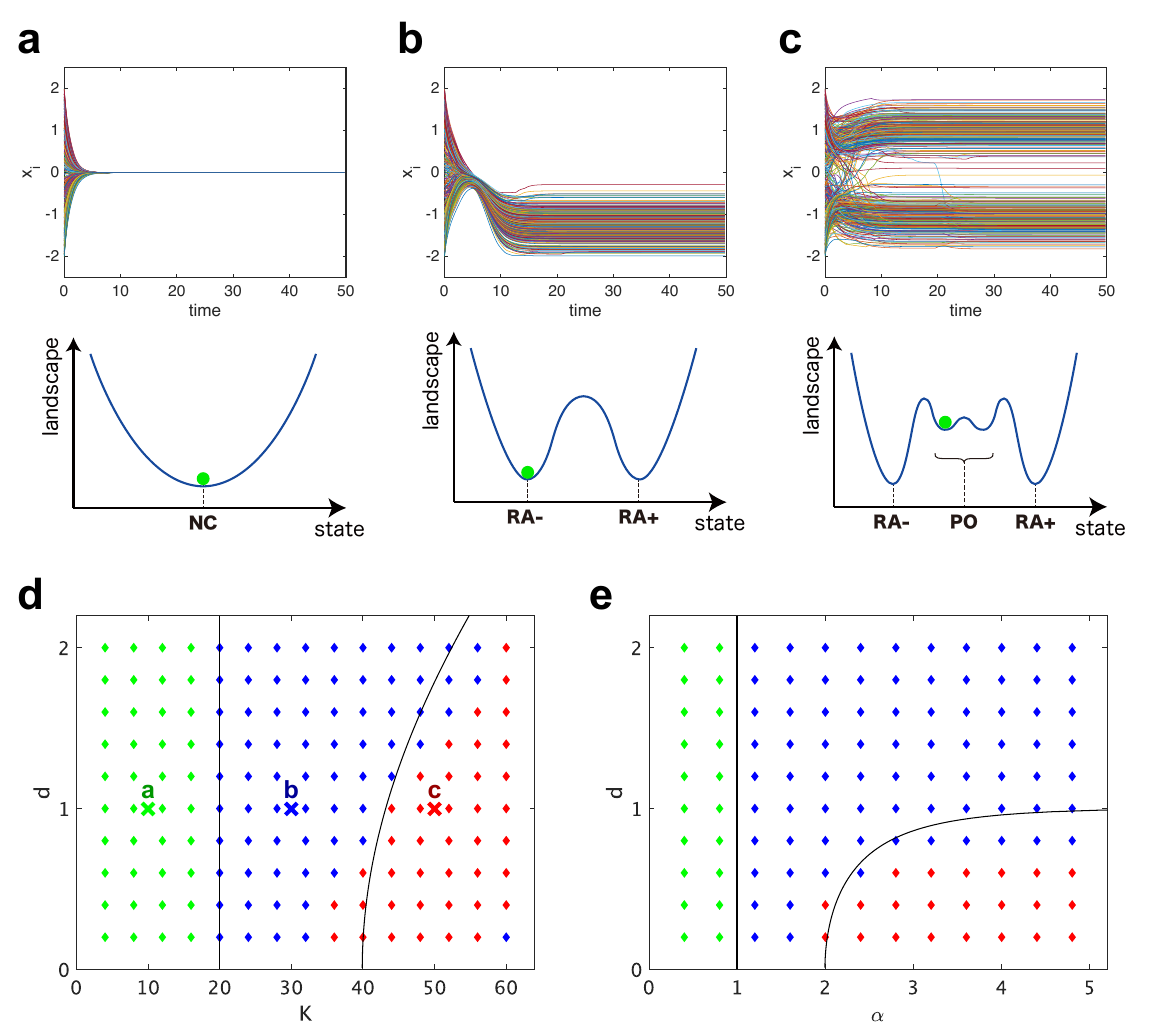}
    \caption{Three typical opinion states and the phase diagram of the opinion dynamics model. \\ 
    (a)-(c) Time series of the individual users' opinions $\{ x_i \}$ ($i= 1, 2, \cdots , N$) (top) and the stability landscape (bottom) approaching three states: 
    (a) Neutral Consensus (NC): All opinions $x_i$ converge to zero: $x_i= 0$, 
    (b) RAdicalization (RA): All opinions are distributed with the same sign, and 
    (c) POlarization (PO): The opinions split into two groups which localize in different signs. 
    The interaction network $A_{ij}$ is given by the Erd\"{o}s-R\'{e}nyi random graph with the number of nodes $N= 400$ and the average degree $\langle k \rangle= 20$. 
    The parameters are set as $\alpha = 1.0$ and $d=1.0$. Coupling strength is set as (a)$K=10$, (b) $K=30$ and (c) $K=50$. Opinions $x_i$ are initially distributed uniformly in $x_i \in [-2, 2]$. \\
    (d) and (e) Phase diagrams for three states. Green and blue diamonds represent NC and RA, respectively. Red diamonds represents parameter region where both PO and RA states are observed depending on the initial state of $\{x_i\}$. Black curves are theoretical predictions of boundaries for the three phases. 
    Crosses in panel (d) represent parameter values used in panels (a)-(c). 
    Parameters are set as $\alpha= 1$ for (d) and $K= 20$ for (e). 
    }
    \label{fig:Phase_Diagram}
\end{figure*}

We first study the collective opinion of the mathematical model (\ref{eq_OD}) on a random network (Erd\"{o}s-R\'{e}nyi network). The Erd\"{o}s-R\'{e}nyi network is a simple random graph model in which each pair of users is connected independently with probability $p$.
As illustrated in Fig.\ref{fig:Phase_Diagram}a-c, the model exhibits three distinct regimes: (a) Neutral Consensus (NC), where the opinions of all users converge towards the neutral opinion $x_i = 0$, (b) Radicalization (RA), where the majority of users hold positive or negative opinions, and (c) Polarization (PO), where two opinion groups (positive and negative) are formed. 
In RA and PO states, we observe a dispersion of the asymptotic opinion values $x_i$, which may be explained by the heterogeneity of the node degrees, and the fact that users with a large number of links tend to hold more extreme opinions (large values of $|x_i|$) due to their more frequent interactions Supplementary Fig. 1). 

Let us investigate numerically the conditions for each regime to emerge, focusing on the role of three parameters: the coupling strength $K$ between users, the user's tolerance $d$, and the controversy of the issue $\alpha$.
We define that the collective opinion is Neutral Consensus (NC) if all users' opinions converged to $0$, RAdicalization (RA) if the distribution of users' opinions was concentrated in one peak or one sign (positive or negative), and POlarization (PO) if the distribution of users' opinions was in two peaks, positive and negative (see Method for definition). 
For each parameter, we ran simulations from 100 different initial conditions $x_i(0)$ to examine realizable collective opinion state.

NC (Fig.~\ref{fig:Phase_Diagram}a) was observed when the coupling strength $K$ was small and the controversy $\alpha$ was small (Fig.~\ref{fig:Phase_Diagram}d and e). Theoretically, we can derive the condition for the NC state to be realized as $K \alpha < p^{-1}$, where $p$ is the link probability of the Erd\"{o}s-R\'{e}nyi network (see Methods for derivation). 
This equation explains the region in which the neutral state is realized, which was obtained by numerical calculation. 
In this parameter regime, the NC is the only realizable state; the RA and PO are not realizable. 
This result can be interpreted using the stability landscape, where the opinions of the users are interpreted as a specific level, and the temporal evolution of the system is seen as a process in which the opinion state of users converges to a state of minimal level. This opinion dynamics system can be interpreted as a system with a single valley that corresponds to the NC state (Fig.~\ref{fig:Phase_Diagram}a, bottom). 
This result is consistent with the fact that, when $K$ or $\alpha$ is sufficiently small, opinion dynamics (\ref{eq_OD}) can be approximated by $\dot{x}_i= -x_i$ and this equation has only neutral state solutions. 
In addition, the user's tolerance $d$ does not affect the feasibility of the NC state. This is because in the NC state, $x_j- x_i= 0$ and $g(|x_i-x_j|)= 1$ is thus independent of $d$. 

RA (Fig.~\ref{fig:Phase_Diagram}b) was observed when the coupling strength $K$ or the controversy $\alpha$ increased (Figs.~\ref{fig:Phase_Diagram}d and e). Theoretically, the conditions under which the RA state is possible are given by $K \alpha > p^{-1}$ (see Method), and this equation can explain the region in which the RA state is realized, in agreement with numerical experiments. 
In this parameter region, we observed the RA$+$ state, in which opinions are concentrated on the positive side, and the RA$-$ state, in which opinions are concentrated on the negative side. Which of the RA$+$ and RA$-$ states is realized depends on the initial state of opinions. From the stability landscape perspective, the opinion dynamics can be interpreted as a system with double valleys that correspond to the RA$+$ and RA$-$ states (Fig.~\ref{fig:Phase_Diagram}b, bottom).  

Furthermore, when the coupling strength $K$ or the controversy $\alpha$ is large and the tolerance $d$ is small, a polarized state (Fig.~\ref{fig:Phase_Diagram}c) can be observed (Figs.~\ref{fig:Phase_Diagram}d and e). In this parameter region, two opinion states, RA and PO, were observed depending on the initial state. 
Theoretically, the condition under which the PO state is feasible is derived as $d < K p \tanh \left( \alpha d/2 \right)$ (see Method for derivation). 
This formula can explain the realization domain of the PO state obtained numerically. In particular, it can be understood from this formula that the controversiality $\alpha$ needs to be large, and the tolerance $d$ needs to be small. 
It is important to note that both RA states and various PO states can be realized within this parameter space. In this context, the PO states refers to multiple states in which user groups with positive or negative opinions differ. From the stability landscape perspective, the opinion dynamics can be interpreted as a system with multiple valleys that correspond to the RA and PO states (Fig.~\ref{fig:Phase_Diagram}c, bottom).

\begin{figure}[t]
    \centering
    \includegraphics[width= 1.0\hsize]{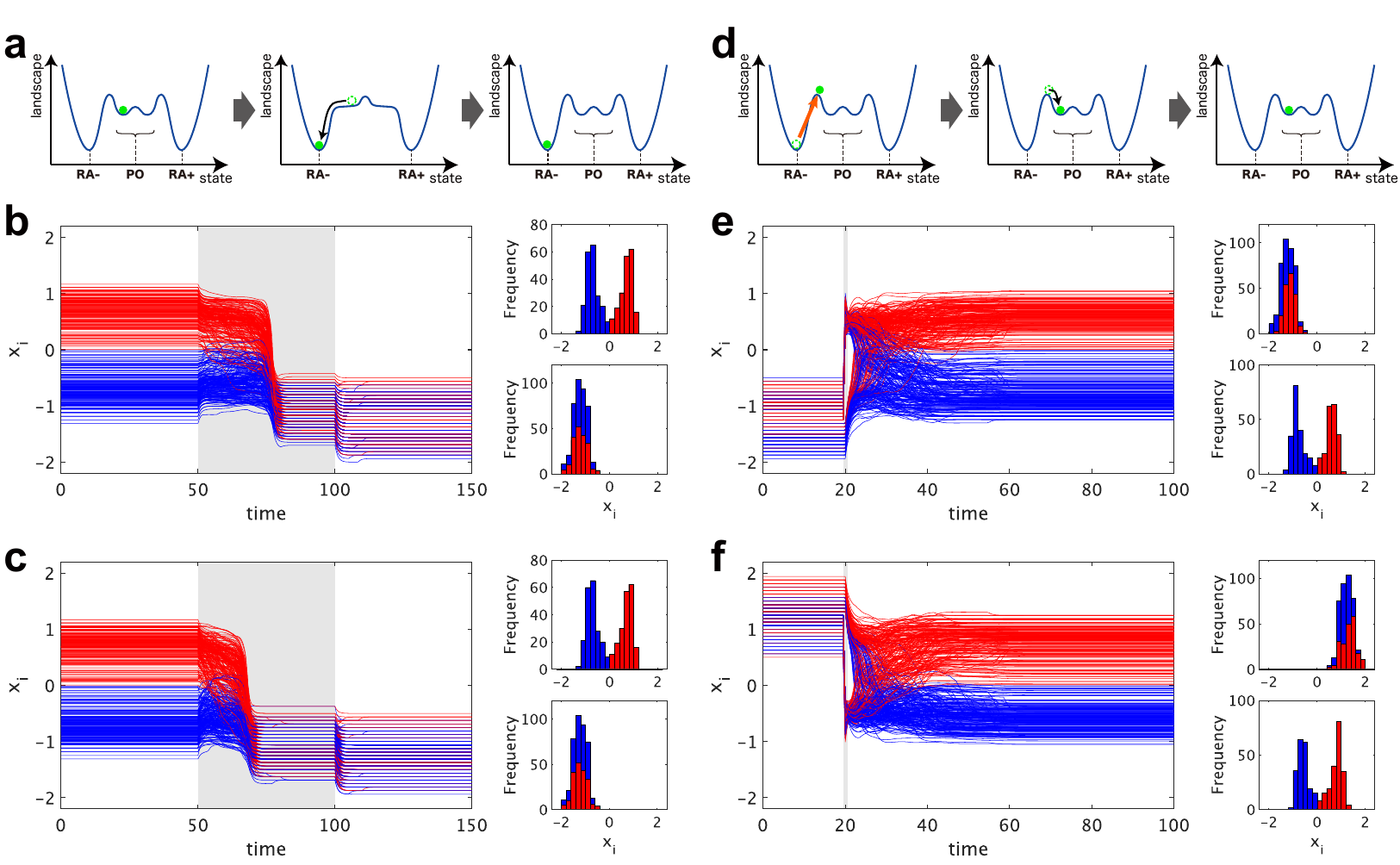}
    \caption{
    Manipulating the opinion state of users.  \\
    (a) Stability landscape interpretation to induce the opinion state via social network intervention: Polarization (PO) to Radicalization (RA).  
    Left: Initially, the opinions are in the PO state, while the RA state is also stably attainable. 
    Center: An intervention in the social network has the effect of altering the stability landscape, thereby allowing only the RA state.
    Right: The opinion state remains RA even after the intervention. \\
    (b), (c): Two types of interventions in the social network: the decrease in the coupling strength $K$ (b) and the removal of the links (c). 
    Left: Time series of individual opinions $x_i$. 
    Right: Histograms of the opinions before (top: $t = 0$) and after (bottom: $t= 150$) the intervention. The shaded areas represent the intervention period ($t \in [50, 100]$). 
    Note that users with positive (negative) opinions before the intervention are shown in red (blue). The coupling strength and the links were reduced in 20 \%.   \\  
    (d) Stability landscape interpretation to induce the opinion state by external driving forces (e.g., propaganda): Radicalization (RA) to polarization (PO).
    Left: Initially, opinions are in the RA state.
    Center: External driving forces cause opinions to surpass the barrier and move closer to the PO state. 
    Right: The opinion state remains PO after the intervention.  \\
    (e), (f): Induction from the RA state, in which most users' opinions are either positive (e) or negative (f), to the PO state. Note that users with positive (negative) opinions at time $t=100$ are shown in red (blue). \\    
    All numerical simulations are done on Erd\"{o}s-R\'{e}nyi network of size $N= 400$ and mean degree $\langle k \rangle= 20$. Parameters are set as $K = 20$, $\alpha= 3$ and $d= 0.7$. }    
    \label{fig:Opinion_Manip}
\end{figure}

\subsection*{Large-scale opinion manipulation }

Results of the previous section (Figs.~\ref{fig:Phase_Diagram}a-c) suggest that the stability landscape perspective\cite{scheffer2001catastrophic,waddington2014strategy,steffen2018trajectories} provides an interpretation on the opinion dynamics in the NC, RA, and PO parameter regime. 
In this subsection, we explore the possibility of manipulating the opinion state through strategic interventions from the perspective of the stability landscape. 
We have just examined the relationship between the parameters $(K, d, \alpha)$ and the asymptotic opinion state (Fig.~\ref{fig:Phase_Diagram}), where $K$ is the coupling strength of the social network, and $d$ and $\alpha$ are parameters that characterize the user and the issue. 
In principle, an SNS provider has the possibility to perform interventions in social networks \cite{bond201261}.  
In contrast, modifying the user properties would be challenging, as it would require altering the behavior of individual users. 
For this reason, we consider the problem of opinion manipulation through a targeted modification of the effective coupling strength, including the social network and the coupling strength $K$, even though a similar analysis could be done when modifying the  parameters $d$ and $\alpha$.

In this study, we address the question of 
whether social network interventions can manipulate users' overall opinions, such as  polarization (PO) and radicalization (RA). 
Specifically, we consider two cases for the manipulation of the opinion states: 1) the transition from PO to RA, and 2) the transition from the RA to PO.  
First, let us consider the manipulation from the PO state to the RA state (Figs.~\ref{fig:Opinion_Manip}a-c). 
We observe that the RA state can also be formed in the parameter domain where the PO state is realized (Figs.~\ref{fig:Phase_Diagram}d and e). This implies that for the parameter regions where PO is possible, the stability landscape has multiple valleys corresponding to PO and RA. 
Additionally, in parameter regions where RA can be formed but PO cannot, the stability landscape contains only the valleys corresponding to RA$+$ and RA$-$. 
Therefore, it is expected that the PO state can be transitioned to the RA state by temporarily adjusting the parameters in which only the RA valleys exists, and then returning the parameters to their original values (Fig.~\ref{fig:Opinion_Manip}a). The opinion state will remain the RA state, because both PO and RA states should be stable at the original parameters. 
To verify this hypothesis, we have conducted a series of numerical experiments. As illustrated in the phase diagram (Figs.~\ref{fig:Phase_Diagram}d and e), it is necessary to reduce the effective coupling strength to achieve the parameter region where only the RA state is formed. 
We perform two types of interventions: reducing the coupling strength $K$ and removing some of the links. 
Note that it is possible to implement a reduction in the coupling strength $K$ by concealing specific posts directed at followers. Similarly, it is also possible to remove the links of a social network by removing users from the follower list.  
Figures~\ref{fig:Opinion_Manip} (b) and (c) demonstrate that these two operations can lead to the RA state, where most users have positive or negative opinions.  
Then, we address the second question: Which user information is necessary for opinion manipulation? 
Transitioning from the PO state to the RA state does not require user information. Opinion manipulation can be achieved by reducing the overall coupling strength or by removing links. However, this strategy does not allow for the manipulation of user opinion, i.e., it cannot specify whether the majority of users agree or disagree with an issue. The RA state's opinion is contingent on the opinions of the users and the social network at the time of manipulation.

Next, we consider the feasibility of manipulating the opinion state from RA to PO. 
Figures~\ref{fig:Phase_Diagram} (d) and (e) suggest that there is no parameter region where only the PO state can be realized. Therefore, it seems implausible to induce the opinion state from RA to PO through social network intervention alone. For example, it is not possible to manipulate the opinion state from RA to PO by temporarily increasing the effective coupling strength between users (Supplementary Fig. 2). 
These results indicate that a social network intervention is inadequate to shift the opinion state from RA to PO. 
For this reason, we consider an alternative strategy by using external driving forces, such as propaganda, to surpass the barrier in the stability landscape (Fig.~\ref{fig:Opinion_Manip}d). 
We consider the RA$-$ state, in which most users disagree on the issue ($x_i<0$), in the initial condition. In our experiments, the opinion state is shifted closer to the PO state by applying an external force, $\Delta x$, to users whose opinion $x_i$ is smaller than a certain value: $x_i \rightarrow x_i + \Delta x$. This operation successfully changes the opinion state from RA$-$ to PO (Fig.~\ref{fig:Opinion_Manip}e). In addition, the strategy effectively shifts the opinion state from PO$+$ (where most users are opposed to the issue: $x_i< 0$) to PO (Fig.~\ref{fig:Opinion_Manip}f).
Our simulation results thus demonstrate that it is possible to manipulate opinion states from RA to PO by applying strong external forces (e.g., propaganda) to users. 
\vspace{1cm}
\clearpage

\begin{figure}[h]
    \centering
    \includegraphics[width=1.0\hsize]{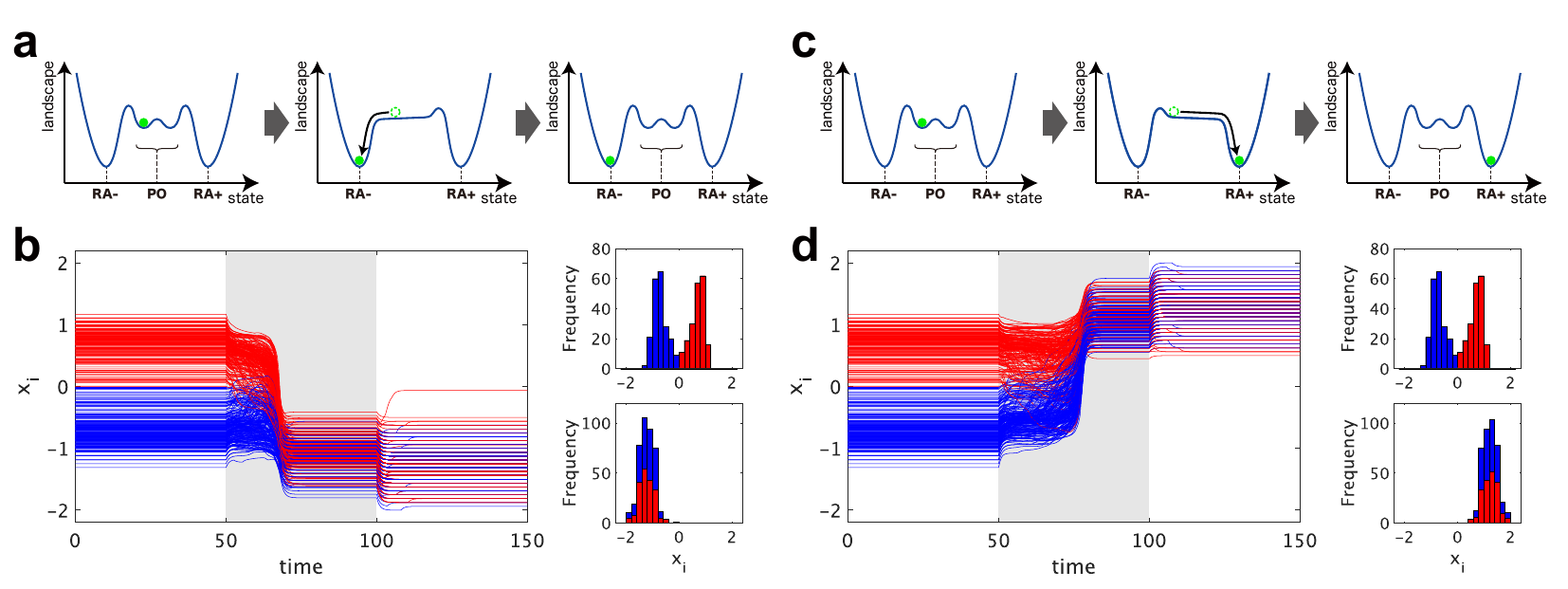}
    \caption{    
    Manipulation toward a targeted opinion state through a selective intervention in a social network. \\
    (a) Stability landscape interpretation to manipulate the opinion state to a targeted state via social network intervention: Polarization (PO) to Radicalization with negative opinions (RA$-$).  
    Left: Initially, the opinions are in the PO state, while the RA$-$ state is also stably attainable. 
    Center: A selective intervention in the social network has the effect of 
    breaking the symmetry in the stability landscape, thereby allowing that the RA$-$ state is more realizable than the RA$+$ state.
    Right: The opinion state remains RA$-$ even after the intervention. \\
    (b) A radicalized state of users with negative opinions (RA$-$ state) can be induced by $20\%$ of the coupling strengths between the users with positive opinions ($x_i > 0$) and $10\%$ of the coupling strengths of the other links. \\
    (c) Stability landscape to manipulate the opinion state to a targeted state: Polarization (PO) to Radicalization with positive opinions (RA$+$). 
    Left: Initially, opinions are in the RA state.
    Center: A selective intervention in the social network has the effect of 
    breaking the symmetry in the stability landscape, thereby allowing that the RA$+$ state is more realizable than the RA$-$ state. 
    Right: The opinion state remains RA$+$ even after the intervention. \\
    (d) A radicalized state of users with positive opinions (RA$+$ state) can be induced by randomly removing $20\%$ of the links between the users with negative opinions ($x_i < 0$) and $10\%$ of the coupling strengths of the other links.
    Note that the shaded area in panels (b) and (d) indicates the intervention period ($50<t<100$). The users with positive (negative) opinions before the intervention are shown in red (blue). 
    }
    \label{fig:Targeted_Manip}
\end{figure}

\begin{figure}[h]
    \centering
    \includegraphics[width=0.85\hsize]{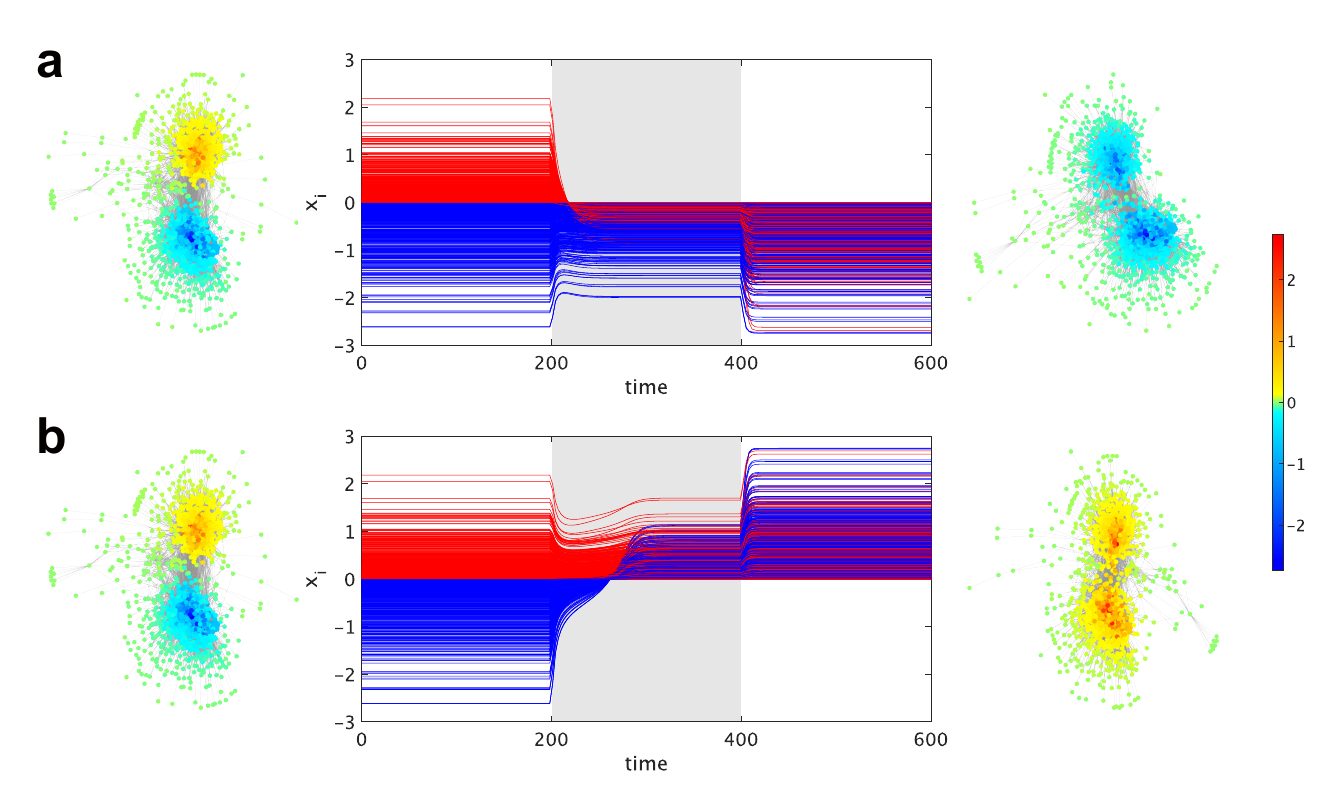}
    \caption{
    Manipulating the opinion state on a Twitter network~\cite{garimella2018quantifying}. \\
    (a) A radicalized state of users with negative opinions (RA$-$ state) can be induced by $20\%$ of the coupling strengths between the users with positive opinions ($x_i > 0$) and $10\%$ of the coupling strengths of the other links. 
    (b) A radicalized state of users with positive opinions (RA$+$ state) can be induced by randomly removing $20\%$ of the links between the users with negative opinions ($x_i < 0$) and $10\%$ of the coupling strengths of the other links. Note that the shaded area indicates the intervention period ($200< t <400$). 
    The users with positive (negative) opinions before the intervention are shown in red (blue). 
    The giant component of effective social networks whose adjacency matrix is given by $A_{ij} g(|x_i- x_j|; d)$ before (left: $t=0$) and after (right: $t=600$) intervention are visualized using force layout algorithm in MATLAB. The color of a node represents the opinion of each user (red: $0 < x$, blue: $x< 0$). Parameters are set as $K= 400$, $\alpha= 3.0$, and $d= 2.5$.  
    }
    \label{fig:Fig5}
\end{figure}

As demonstrated in our previous analysis (Figs.~\ref{fig:Opinion_Manip}a--c), interventions within a social network can facilitate transitions in opinion states from polarization (PO) to radicalization (RA). 
However, this strategy does not guarantee that the majority of users will be steered toward a specific stance -- either agreement (PO$+$) or disagreement (PO$-$)-- on the issue. 
To address this limitation, we address the question of whether the social network intervention can guide even the agree (RA$+$) or disagree (RA$-$) state by leveraging the opinion state of individual users. 
To this end, we consider three opinion states: PO, RA$+$, and RA$-$ (Fig.~\ref{fig:Targeted_Manip}a) from the stability landscape perspective. 
Suppose the users' opinions, ${\bf x}^+= (x^+_1, x^+_2, \cdots , x^+_N)$, represent a solution corresponding to the RA$+$ state. 
By inverting the sign of all users' opinions, we obtain ${\bf x}^- := -{\bf x}^+$, which constitutes a solution in the RA$-$ state. This is due to the fact that the controversy function $h(x_j)$ is an odd function. 
Moreover, it can be shown that the linear stability of the two states (${\bf x}^+$ and ${\bf x}^-$) is equivalent. Consequently, the system exhibits symmetry, meaning that the shapes of the RA$+$ and RA$-$ valleys in the stability landscape are similar (Fig.~\ref{fig:Targeted_Manip}a). 
To selectively induce either the RA$+$ or RA$-$ state, it is necessary to break the symmetry. 

A potential strategy to break the symmetry in polarized states is to reduce the effective coupling within an opinion community. For instance, the reduction in the coupling strengths among users with positive opinions ($x_i > 0$) could lead to the fragmentation of the positive-opinion group, thereby facilitating a transition to the RA$-$ state. 
This occurs because the reduced coupling inhibits mutual reinforcement among users with similar views, which makes it difficult to maintain the opinion community. 
We temporarily reduced the coupling strengths within the user community that agreed with an issue ($x_i > 0$) in the PO state by 20 \% and those of other links by 10 \%. 
This operation successfully transitioned the opinion state from PO to RA$-$, where the majority of users expressed disagreement with the issue (Fig.~\ref{fig:Targeted_Manip}a).
Similarly, we can induce an opinion state from PO to RA$+$ (Fig.~\ref{fig:Targeted_Manip}b) by selectively reducing the coupling strength of links within a community of users with opposing opinions ($x_i < 0$). 
These results (Fig.~\ref{fig:Targeted_Manip}) demonstrate that a social network intervention can selectively transform a polarized state into a radicalized one, in which the majority is either in agreement (RA$+$) or disagreement (RA$-$). This manipulation requires identifying the stance of some users -- whether they agree or disagree.
It is possible to infer users' stances on the issue from the content of posts and reposts \cite{conover2011political,jiang2023retweet}. Therefore, this manipulation is feasible for issues that attract considerable attention.

We have demonstrated that interventions in social networks can shift the opinion state from polarization (PO) to radicalization (RA) (Fig.~\ref{fig:Opinion_Manip}). 
Furthermore, it is possible to transition from the PO state to a targeted RA state, where most users express agreement or disagreement on an issue (Fig.~\ref{fig:Targeted_Manip}).
We verify these results using real network data from Twitter (X). We simulated the opinion dynamics model using the follow-follower network with the hashtag (\#russia\_march in Garimella et al. 2018~\cite{garimella2018quantifying}). 
A polarized opinion state is simulated by setting the initial state of each user group as either "agree" ($x_i > 0$) or "disagree" ($x_i < 0$) for a given issue (See Method for details). 
We temporarily decrease the coupling strengths between users who agreed ($x_i > 0$) by 20 \%, those of other links by 10 \%, and then return them to their original values. Consequently, we effectively changed the overall opinion state from PO to RA$-$ (Fig.~\ref{fig:Fig5}, Top).
Similarly, we are able to manipulate the opinion state from PO to RA$+$ (Fig.~\ref{fig:Fig5}, bottom) by temporarily decreasing the coupling strength between users who disagree ($x_i < 0$) by 20 \% and those of other links by 10 \% on the issue.

\section*{Discussion}

In this study, we proposed and analyzed a mathematical model of opinion dynamics capable of reproducing key phenomena observed in online social media, namely opinion polarization and radicalization. Through theoretical and numerical analyses, we identified three distinct steady-state regimes: Neutral Consensus (NC), where most individuals maintain neutral opinions; Radicalization (RA), where the majority adopt either strongly supportive (RA$+$) or strongly opposing (RA$-$) views; and Polarization (PO), characterized by the coexistence of two opposing opinion groups. Within the scope of our model, the emergence of each regime depends on system parameters such as interaction strength $K$, issue controversy $\alpha$, user tolerance $d$, and initial opinion distribution. We further explored the potential for influencing collective opinion through external interventions. Our results show that targeted structural interventions in the network can drive transitions from the PO state to either favorable (RA$+$) or unfavorable (RA$-$) radicalized states (Figs.~\ref{fig:Opinion_Manip}a--c, ~\ref{fig:Targeted_Manip}). Remarkably, this approach remains effective even when applied to real-world network data, as we demonstrated in a Twitter-based interaction graph (Fig.~\ref{fig:Fig5}). In contrast, reversing this transition, from RA to PO, proved infeasible without the addition of a collective external influence, such as propaganda targeting a large portion of the population.

While our model captures key mechanisms behind opinion polarization and radicalization, it is based on several simplifying assumptions. First, we considered opinions on a single, isolated issue. Although our findings hold for multiple independent issues, future work should explore how interdependencies between issues influence opinion dynamics \cite{tian2025matrix}, particularly in environments where topics reinforce or conflict with one another. Second, we assumed homogeneous users who conform to the opinions of their peers. However, empirical studies suggest that some individuals may respond antagonistically to opposing views, exhibiting a repulsive response to disagreement~\cite{bail2018exposure,sabin2020pull}. Incorporating heterogeneous user behaviors into the model could yield a richer understanding of real-world dynamics. Third, for analytical tractability, we assumed a static interaction network representing continuous interpersonal influence. In practice, online interactions are event-driven and temporally structured. Although modeling interactions as a Poisson process may not drastically alter the qualitative outcomes, exploring non-Poissonian features, such as bursty human communication patterns, could provide deeper insight into their role in opinion evolution~\cite{masuda2016guide,aoki2016input,zarei2024bursts}.

Beyond theoretical contributions, our findings raise ethical considerations. Network interventions, such as modifying interaction structures, offer a powerful means of steering collective opinion. Unlike overt propaganda, these interventions can be subtle and less perceptible to users, making them both effective and potentially concerning. While such strategies might be employed with benevolent intent (e.g., to reduce polarization), they can also be misused for manipulation. Thus, great caution is required when designing or deploying mechanisms that alter social connectivity.
Our results underscore the importance of developing mathematical frameworks that not only reproduce observed patterns of polarization and radicalization but also reveal the mechanisms leading to the emergence of collective phenomena. A rigorous understanding of these dynamics is essential for assessing the risks and unintended consequences of large-scale interventions aimed at shaping public discourse.
Ultimately, our study highlights the delicate balance between promoting healthy online interaction and avoiding interventions that may inadvertently deepen societal divides. The structure of digital platforms profoundly shapes how opinions spread, polarize, or converge. Careful design, informed by theoretical insights, is crucial to fostering resilient and inclusive public debate in an increasingly networked world.

\vspace{1cm}

\section*{Methods}

\subsection*{Opinion dynamics model}  

We consider a simple mathematical model of opinion dynamics in a social network. 
This model consists of $N$ individuals (users) who exchange opinions about a binary (yes or no) issue. 
The $i$-th user is characterized by its opinion $x_i \in (-\infty, \infty)$, which represents the user's stance on the issue, indicating whether the user is for ($x_i > 0$) or against ($x_i < 0$) the issue. The magnitude of $x_i$ represents the strength of the opinion, with a large $|x_i|$ indicating a strong agreement or disagreement with the issue.   
The opinion of users ($i= 1, 2, ..., N$) is described by a differential equation: 
\begin{equation}
	\frac{\textrm d x_i}{\textrm d t} = -x_i +\frac{K}{N} \sum_{j=1}^N A_{ij} g(|x_i - x_j|; d) h(x_j; \alpha), 
	\label{eq_OD_Method}
\end{equation}
where $K$ is the strength of social interaction between users and $A_{ij}$ are the entries of the adjacency matrix of the social network. 
The effect of social interaction on individual opinions is described by $g(|x_i - x_j|; d) h(x_j; \alpha)$. 
The homophily function $g(|x_i - x_j|; d)$ describes the tendency of individuals to interact with people with similar opinions. We focus on a specific case inspired by the bounded confidence model~\cite{deffuant2000mixing,gargiulo2012influence,axelrod2021preventing,sasahara2021social} (see Supplementary Note for the comparison to other opinion dynamics model), $g(|x_i - x_j|; d) = \theta(d - |x_i - x_j|)$, where $\theta(x)$ is the Heaviside function, i.e. $\theta(x)=1$ when $x\geq 0$ and $\theta(x)=0$ when $x<0$, and $d$ quantifies the opinion tolerance of a user: the user $j$ whose opinion disagreement is greater than $d$ does not influence the opinion of the user $i$. 
As we show in the Supplementary Information, the results are qualitatively similar for other choices of functions $g$ (Supplementary Figs. 3 and 4). 
The controversy function $h(x_j; \alpha)$ describes the social reinforcement on the issue, and takes the form $h(x_j; \alpha) = \tanh (\alpha x_j)$~\cite{baumann2020modeling}, where $\alpha$ represents the level of controversy surrounding the issue. 
In the absence of social interaction, the opinions of all individuals converge to a neutral state (i.e., $x_i= 0$). 
In addition to the theoretical analysis (see subsections "Mean field equations" and "Stability analysis of opinion states"), the opinion dynamics model (\ref{eq_OD_Method}) is solved numerically using the Runge-Kutta method with a step size of $1.0 \times 10^{-4}$.

\subsection*{Mean field equations}  
An all-to-all connected opinion dynamics model that corresponds to a random network (Erd\"{o}s-R\'{e}nyi network: Eq.(\ref{eq_OD_Method}) ) can be approximated by replacing each component of the adjacency matrix with its expected value, $A_{ij} \approx \frac{ \langle k \rangle }{N-1}$,    
\begin{equation}
	\frac{\textrm d x_i}{\textrm d t} = -x_i +\frac{\kappa}{N} \sum_{j=1}^N g(|x_i - x_j|; d) h(x_j; \alpha), 
	\label{eq_all2all}
\end{equation}
where $\kappa= \frac{K \langle k \rangle }{N-1}$ is the effective coupling strength, $\langle k \rangle$ is the mean degree of the Erd\"{o}s-R\'{e}nyi network, and $N$ is the number of users. 
As a reminder, the functions $g(|x_i - x_j|)$ and $h(x_j)$ describe homophily in social interactions, i.e., the tendency of individuals to interact with people who share their opinions, and the controversy of an issue, respectively. 

Numerical simulations suggest that this model realizes the neutral consensus (NC) state, the radicalization (RA) states, where the opinion of each user $x_i$ ($i= 1, 2, \cdots, N$) converges to the same value, and the polarization (PO) state, where the opinion of each user converges to either positive or negative values (see Supplementary Fig. 5). In addition, the number of users in each group is approximately equal in the PO state. 
Motivated by the numerical result, we analyze the stability condition of the opinion state when there are two groups with the same opinion ($X$ and $Y$) of the same size. 
Let us consider a user who belongs to the group with opinion $X$. The opinion dynamics model (\ref{eq_all2all}) can be reduced by substituting $X$ to the opinion of the user $i$, and assuming that each half of the other users have opinions $x_j= X$ and $x_j=Y$, respectively: 
\begin{equation}
    \OD{X}{t} = - X + \frac{\kappa}{2} h(X)+ \frac{\kappa}{2}g\left ( \left | X-Y \right|; d \right) h(Y),    
\end{equation}
where we can assume $g(0; d)= 1$ without loss of generality.   
Similarly, we consider a user in the group with opinion $Y$. Substitution of $Y$ for the opinion of user $i$ in the opinion dynamics model (\ref{eq_all2all}) yields
\begin{align}
    \OD{Y}{t} = - Y + \frac{\kappa}{2}g\left ( \left | X-Y \right |; d \right ) h(X)+ \frac{\kappa}{2}h(Y).       
\end{align}
By setting the function $g$ as the step function $\theta(x)$ and the function $h$ as the $\tanh$ function, we can derive a mean field equation consisting of two opinion groups
\begin{align}
	\OD{X}{t} = - X + \frac{\kappa}{2} \tanh( \alpha X)+ \frac{\kappa}{2}  \theta \left ( d- \left | X-Y \right |\right ) \tanh( \alpha Y ),	\label{eq_reduced_X}	\\
	\OD{Y}{t} = - Y + \frac{\kappa}{2} \theta \left ( d- \left | X-Y \right |\right ) \tanh( \alpha X)+ \frac{\kappa}{2} \tanh( \alpha Y).  
    \label{eq_reduced_Y}
\end{align}

\subsection*{Stability analysis of opinion states} 
We perform a stability analysis of opinion states using the reduced model described by Eqs. (\ref{eq_reduced_X}) and (\ref{eq_reduced_Y}). 
Three opinion states are considered: neutral consensus (NC), radicalization (RA), and polarization (PO). In a reduced model, the NC state corresponds to $X = Y = 0$; the RA state corresponds to $X = Y$ $(X \neq 0)$ and the PO state corresponds to $X = -Y$ $(X \neq 0)$.

Let us consider the stability condition of NC state. 
NC state ($X=0$, $Y=0$) is always a fixed point, since the derivatives ($\frac{ {\rm d}X}{ {\rm d} t}$ and $\frac{ {\rm d}Y}{ {\rm d}t}$) are zero. 
The fixed point is linearly stable if the Jacobian matrix $J$ at that point 
\begin{align}
    J= \begin{pmatrix}
        -1 + \frac{\alpha \kappa}{2} &  \frac{\alpha \kappa}{2}  \\
        \frac{\alpha \kappa}{2}  &  
        -1 + \frac{\alpha \kappa}{2}
    \end{pmatrix}   \nonumber 
\end{align}
satisfies ${\rm Det}\ J > 0$ and ${\rm Tr}\ J< 0$.
Thus, the stability condition for the NC state is given by $\kappa \alpha < 1$.  
Therefore, the NC state is realized when $\kappa \alpha < 1$, and the RA state or PO state is realized when $\kappa \alpha > 1$. 

For the PO state $(X, Y) = (c, -c)$ ($c>0$) to be realized, it must be a fixed point.
Substituting $(X, Y) = (c, -c)$ into the mean field equation (Eqs. \ref{eq_reduced_X} and \ref{eq_reduced_Y}) yields 
\begin{equation}
    - c + \frac{\kappa}{2} \tanh(\alpha c) 
    \left( 1- \theta(d- 2c) \right) = 0,   
    \label{eq_PO_fixed_point}
\end{equation}
Suppose that $\theta(d- 2c)= 1$, then the above equation (\ref{eq_PO_fixed_point}) indicates that $c= 0$, which is an NC state. Therefore, two opinions in PO states should be sufficiently distinct: $\theta(d- 2c)= 0$, i.e., $2c > d$, and the following equation should be satisfied for PO states
\begin{equation}
    - c + \frac{\kappa}{2} \tanh(\alpha c) = 0,  
    \label{eq_PO_fixed_point2}
\end{equation}
which means that a straight line $y=x$ and a curve $y= \frac{\kappa}{2} \tanh(\alpha x)$ have an intersection point ($x=c$). The derivative of the curve at the intersection point should be smaller than 1: 
\begin{equation}
    \frac{\alpha \kappa}{2} \cosh^{-2}(\alpha c)< 1.    \label{eq_PO_fixed_point3}
\end{equation}
The Jacobian matrix $J$ at the PO state is written as 
\begin{align}
    J= \begin{pmatrix}
        -1 + \frac{\alpha \kappa}{2} \cosh^{-2}(\alpha c) &  0 \\
        0 &  -1 + \frac{\alpha \kappa}{2} \cosh^{-2}(\alpha c)  
    \end{pmatrix}.   \nonumber 
\end{align}
Using Eq. (\ref{eq_PO_fixed_point3}), that PO state is shown to be linearly stable, which satisfies ${\rm det}\ J > 0$ and ${\rm Tr}\ J < 0$. Therefore, PO and RA states can be realized when $ d < \kappa \tanh(\alpha d/2)$. In contrast, only RA state can be realized when $ d > \kappa
\tanh(\alpha d/2)$.

\subsection*{Numerical calculation of phase diagram}
The phase diagram is obtained by simulating the opinion dynamics model (\ref{eq_OD_Method}) and identifying the opinion states based on user opinions. 
First, we simulate the opinion dynamics model (\ref{eq_OD_Method}) by setting the initial state of opinions $\{x_i(0) \}$ $(i = 1,2, \cdots ,N)$ with values drawn uniformly from the interval $[-2, 2]$. 
After the simulation, we obtain the final opinions of all users, denoted as $\{ x_i(T) \}$. 
%
Next, we identify three states of collective opinion based on the distribution of the opinions $\{ x_i(T) \}$. 
a) Neutral Consensus: This state is identified when the sample variance of individual opinions $\{ x_i(T) \}$ is zero. 
For the remaining states---Radicalization and Polarization---we fit the distribution of opinions $\{ x_i(T) \}$ to a Gaussian mixture model. 
The fitting is performed assuming either one or two mixture components, and the model with the minimum Bayesian Information Criterion (BIC) is selected. 
The opinion states are then determined as follows: 
b) Radicalization: This state is identified when the fitted model consists of a single Gaussian distribution or a mixture of Gaussians whose means share the same sign. 
c) Polarization: This state is identified when the fitted model is a mixture of two Gaussian distributions with means of opposite signs (one positive and one negative). 
The MATLAB\copyright\ fitgmdist function was used to fit a mixture of  Gaussian distributions.

According to the theoretical analysis (see Result section), the opinion dynamics model (\ref{eq_OD_Method}) exhibits three phases: 
1) PO \& RA regime: A parameter regime in which the model can exhibit both polarization and radicalization states, 2) RA regime: A parameter regime in which the model can only exhibit the radicalization state. 3) NC regime: A parameter regime in which the model can only exhibit the neutral consensus state.
Therefore, we performed simulations with 100 different initial conditions, and identify the phase as follows: 
1) PO \& RA regime: A parameter set is identified as PO \& RA if polarization was obtained at least once. 
2) RA regime: A parameter set is identified as RA if only radicalization was obtained. 
3) NC regime: A parameter set is identified as NC if only neutral consensus was obtained.  

\bibliography{hata25}

\section*{Acknowledgements}
We thank Takaaki Aoki, Masataka Goto, Ryoich Sasaki, Shigeru Shinomoto, Yuka Takedomi, Takuma Tanaka, Junichi Yamagishi, and JST FOREST researchers for stimulating discussions. 
S.H. acknowledges support from  JSPS KAKENHI 25K15264.
R.L. acknowledges support from the EPSRC Grants EP/V013068/1, EP/V03474X/1, and EP/Y028872/1. 
H.N. acknowledges support from  JSPS KAKENHI 25H01468, 25K03081, and 22H00516. 
R.K. acknowledges support from the JSPS KAKENHI JP21H04571, JP22H03695, and JP23K24950, and JST FOREST JPMJFR232O.

\section*{Author contributions statement}
%
R.K., S.H., and H.N conceived the project. 
S.H and R.K. developed the opinion dynamics model and performed the theoretical analysis.
S.H. performed numerical simulations, analyzed data, and created the figures. 
R.K. and R.L. supervised the project. 
R.K and R.L wrote the draft based on input from S.H. 
All authors edited and approved the manuscript. 

\clearpage
\section*{Supplementary Information}

\subsection*{Supplementary Figures}
\begin{figure}[h]
\begin{center}
    \label{fig:Supp_Fig1}
    \includegraphics[scale=.75]{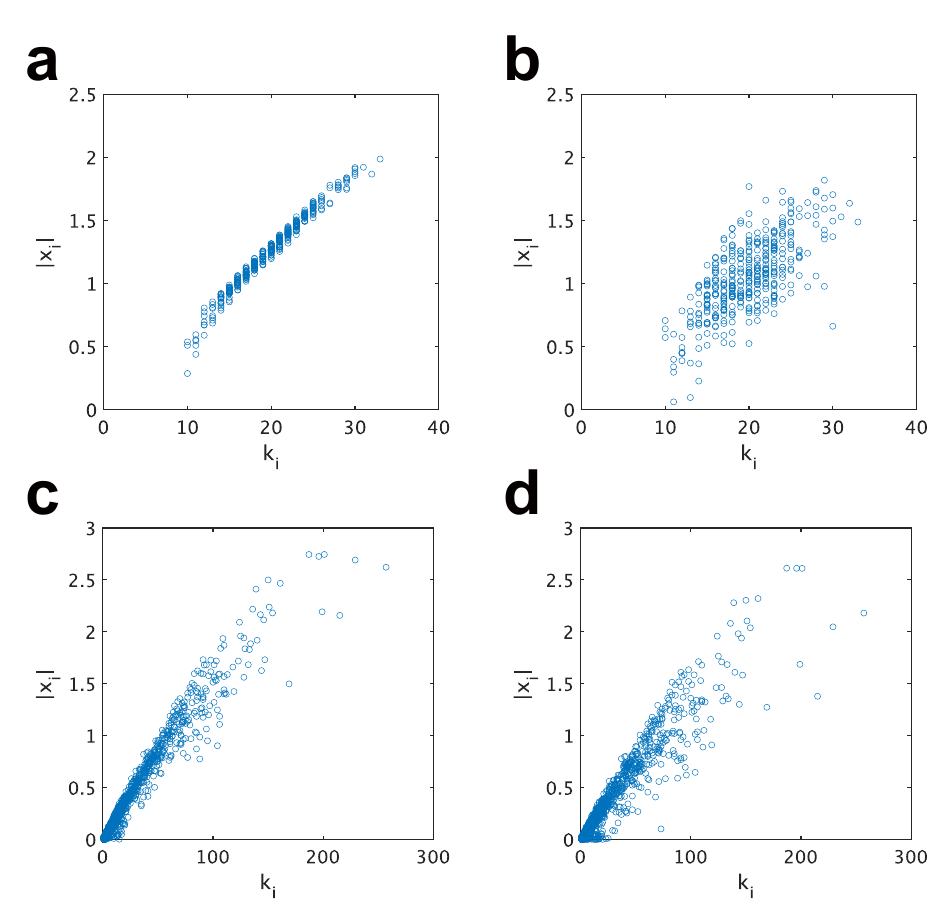} 
    \caption*{ {\bf Supplementary Figure 1}. 
    {\bf Relationship between opinion intensity $|x_i|$ and node degree $k_i$ in the RA and PO states.} \\
    (a) and (b): Results obtained from the Erd\"{o}s-R\'{e}nyi network for the RA (a) and PO (b) states. The parameters and adjacency matrix are the same as those used in Figs.~\ref{fig:Phase_Diagram} (b) and (c) in the main text. \\
    (c) and (d): Results obtained from a Twitter network~\cite{garimella2018quantifying} for the RA (c) and PO (d) states. The parameters and adjacency matrix are the same as those used in Fig.\ref{fig:Fig5} in the main text.
   }   
\end{center}
\end{figure}

\begin{figure}[h]
\begin{center}
    \label{fig:Supp_Fig2}
    \includegraphics[scale=.8]{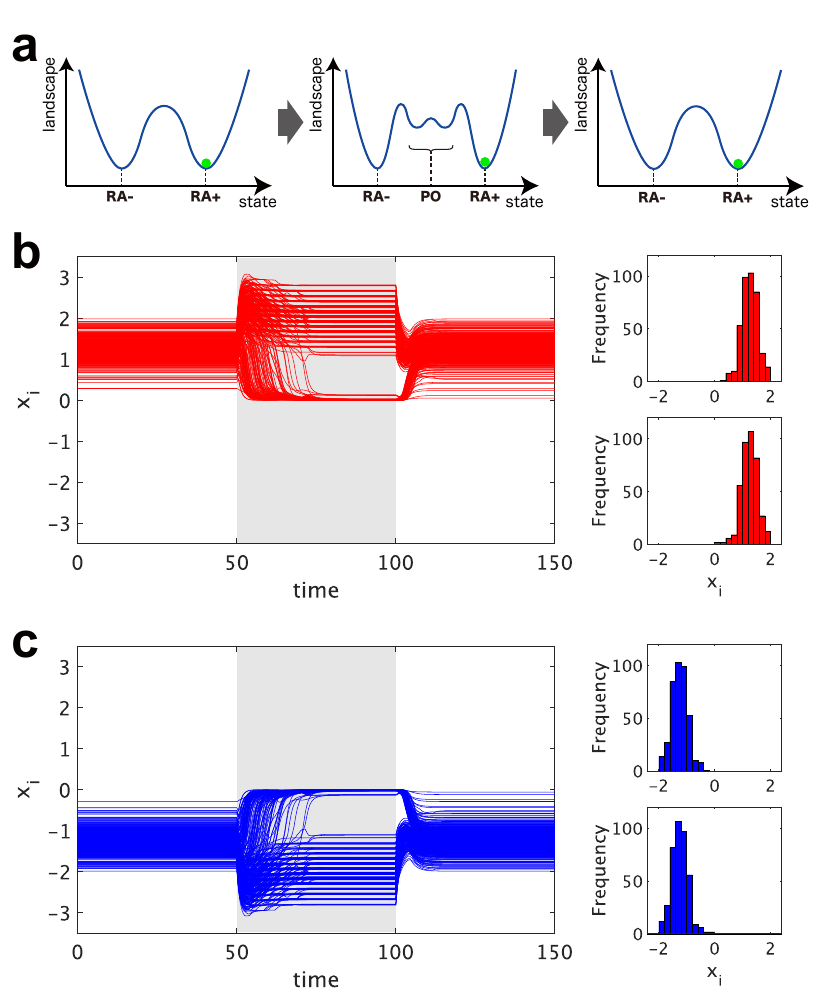} 
    \caption*{ {\bf Supplementary Figure 2}. 
    {\bf  Manipulating the opinion state of users by temporarily increasing the effective coupling strength between users.} \\
    (a) Stability landscape interpretation of the social network intervention. 
    Left: Initially, the opinions are in the RA$+$ state, while the PO state is not realizable.  
    Center: An intervention in the social network alters the stability landscape, thereby allowing both the RA and PO states. However, opinions remain in the RA$+$ state because it is stable. 
    Right: The opinion state remains RA$+$ even after the intervention. \\
    (b), (c): An intervention in the social network, i.e., the temporal increase in the coupling strength $K$, is applied to the RA$+$ (b) and RA$-$ (c) state. 
    Left: Time series of individual opinions  $\{ x_i \}$. 
    Right: Histograms of the opinions before (top: $t = 0$) and after (bottom: $t= 150$) the intervention. 
    The shaded areas represent the intervention period ($t \in [50, 100]$). Note that users with positive (negative) opinions before the intervention are shown in red (blue). The coupling strength before and during the intervention was $K=30$ and $50$, respectively. 
    Other parameters and adjacency matrix are the same as those used in Fig. 3(b) and (c).
   }   
\end{center}
\end{figure}

\begin{figure}[h]
\begin{center}
    \label{fig:Supp_Fig3}
    \includegraphics[scale=.8]{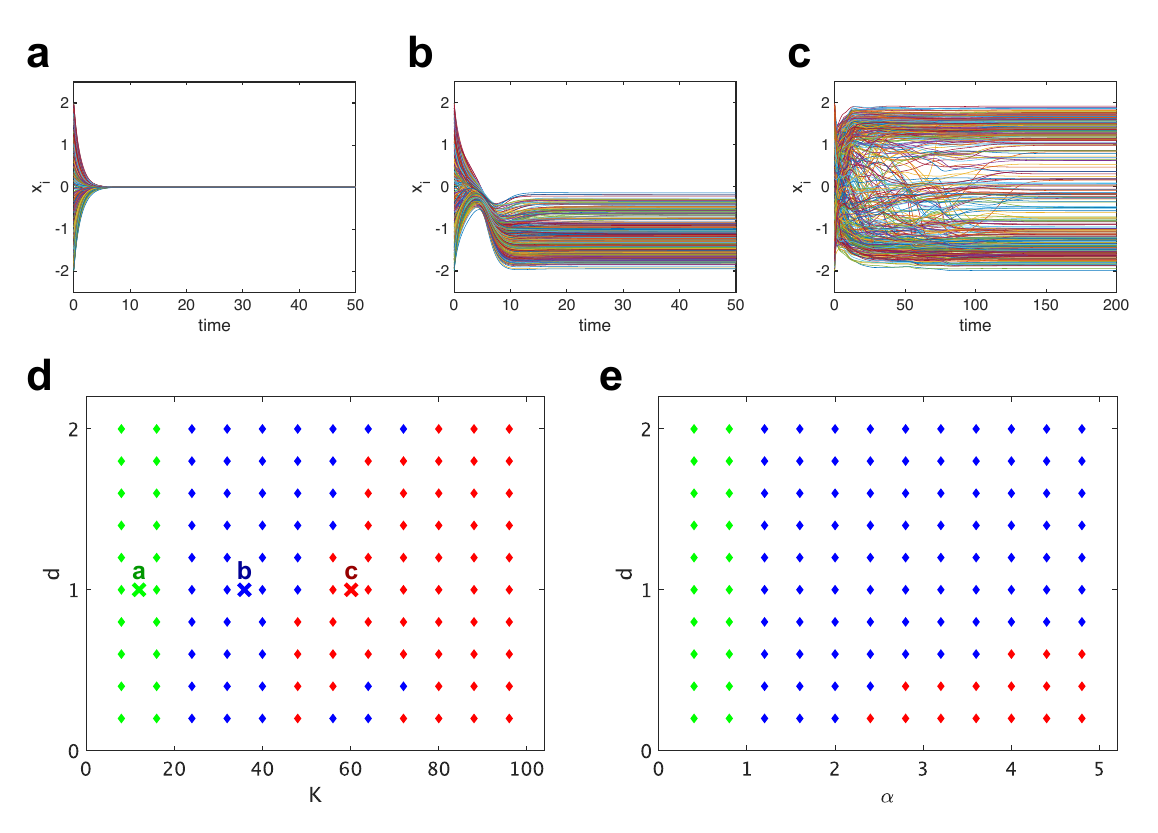} 
    \caption*{ {\bf Supplementary Figure 3}. 
    {\bf Phase diagram of opinion dynamics with a Gaussian-type homophily function.} \\
     (a)-(c): Time series of the individual users’ opinions $\{ x_i \}$ $(i = 1, 2, \dots , N)$, which approaches three states: (a) Neutral Consensus (NC), (b) RAdicalization (RA), and (c) POlarization (PO). 
     The coupling strength was set to $K= 12, 36,$ and $60$, in (a), (b), and (c), respectively. \\
     (d) and (e): Phase diagrams for three states. 
     Green and blue diamonds represent NC and RA, respectively. Red diamonds represents parameter region where both PO and RA states are observed depending on the initial state. Here, we use a Gaussian-type homophily function: 
     $g(|x_i - x_j|; d)= \exp\left[ - \frac{|x_i - x_j|^2}{2d^2}\right]$. 
     The other parameters and the adjacency matrix $A_{ij}$ were the same as in Fig.~\ref{fig:Phase_Diagram}. Crosses in panel (d) represent parameter values used in panels (a)-(c).
   }   
\end{center}
\end{figure}

\begin{figure}[h]
\begin{center}
    \label{fig:Supp_Fig4}
    \includegraphics[scale=.8]{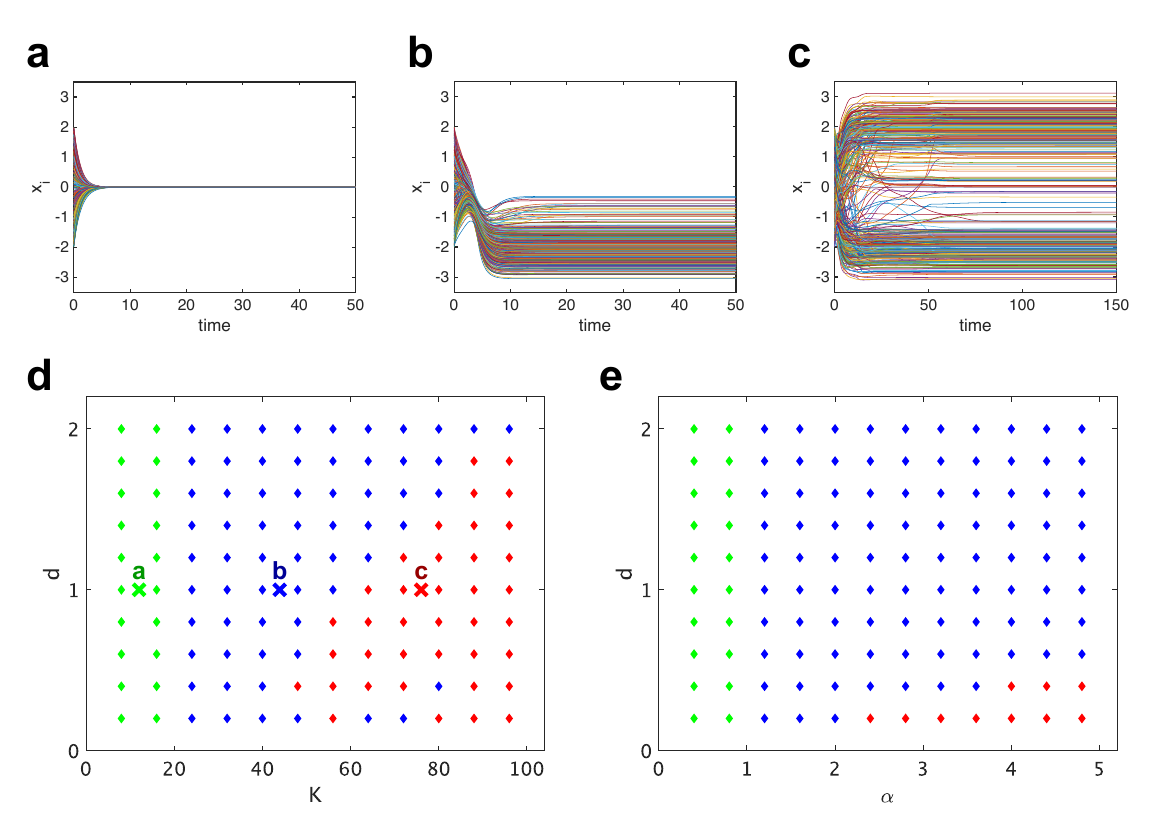} 
    \caption*{ {\bf Supplementary Figure 4}. 
    {\bf Phase diagram of opinion dynamics with a power-law homophily function.} \\
     (a)-(c): Time series of the individual users’ opinions $\{ x_i \}$ $(i = 1, 2, \dots , N)$, which approaches three states: (a) Neutral Consensus (NC), (b) RAdicalization (RA), and (c) POlarization (PO). 
     The coupling strength was set to $K= 12, 44,$ and $76$, in (a), (b), and (c), respectively. \\
     (d) and (e): Phase diagrams for three states. Green and blue diamonds represent NC and RA, respectively. Red diamonds represent parameter region where both PO and RA states are observed depending on the initial state. Here, we use a power-law homophily function: 
     $g(|x_i - x_j|; d)= \frac{\epsilon^{\beta} }{|x_i - x_j|^{\beta} }$ ($|x_i - x_j| \geq d$),   $g(|x_i - x_j|; d)= 1$ ($|x_i - x_j| < d$). Crosses in panel (d) represent parameter values used in panels (a)-(c).
     The other parameters and the adjacency matrix $A_{ij}$ were the same as in Fig.~\ref{fig:Phase_Diagram}. 
   }   
\end{center}
\end{figure}

\begin{figure}[h]
\begin{center}
    \label{fig:Supp_Fig5}
    \includegraphics[scale=.8]{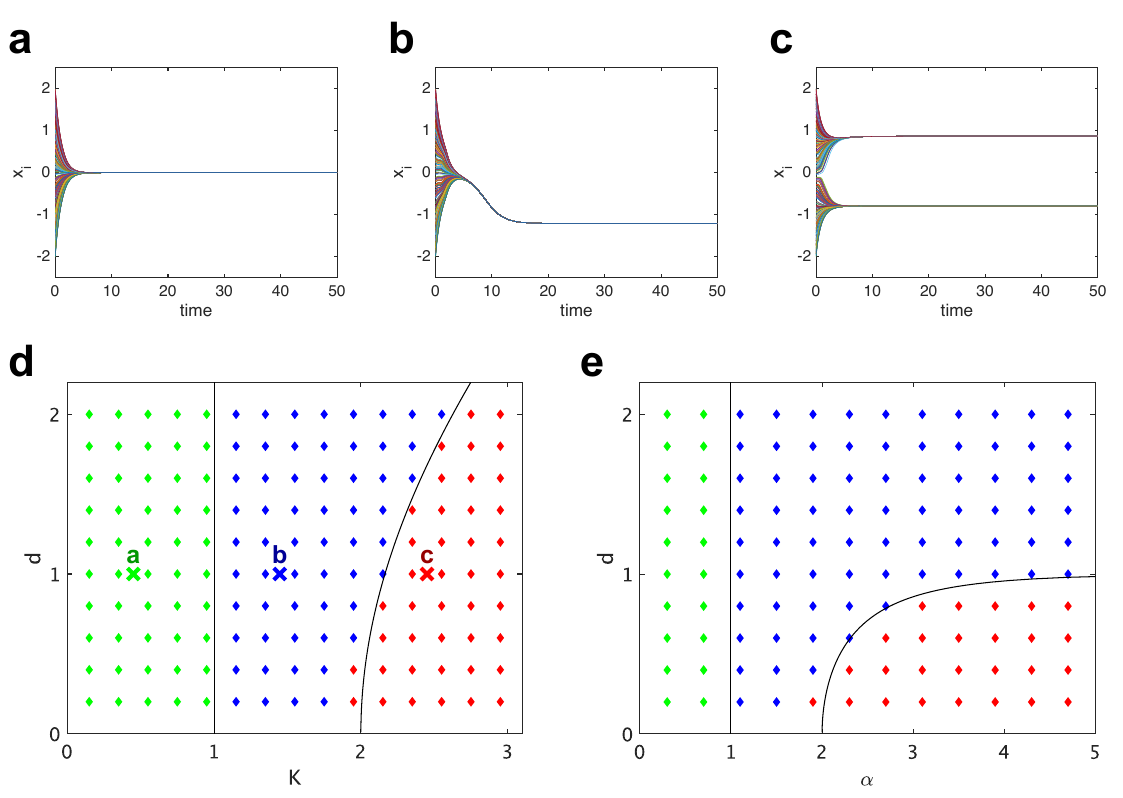} 
    \caption*{ {\bf Supplementary Figure 5}. 
    {\bf Phase diagram of the all-to-all connected opinion dynamics.} \\
     (a)-(c): Time series of the individual users’ opinions $\{ x_i \}$ $(i = 1, 2, \dots , N)$, which approaches three states: (a) Neutral Consensus (NC), (b) RAdicalization (RA), and (c) POlarization (PO). 
     The coupling strength was set to $K= 0.45, 1.45,$ and $2.45$, in (a), (b), and (c), respectively. Other parameters are set as follows: $\alpha= 1$ and $d=1$. \\
     (d) and (e): Phase diagrams for three states.     
     We identify three states of collective opinion based on the distribution of the opinions at time $t=50$:  
     a) NC: This state is identified when the sample mean and variance of individual opinions are zero, b) RA: This state is identified when the sample mean of individual opinions is not zero and the sample variance is zero, and c) PO: This state is identified when the sample variance of individual opinions is not zero. 
     Green and blue diamonds represent NC and RA, respectively. Red diamonds represent parameter region where both PO and RA states are observed depending on the initial state. 
     Here, we examine the all-to-all connected opinion dynamics: $A_{ij}= 1$. Parameters are set as $\alpha= 1$ for (d) and $K=1$ for (e). 
     Crosses in panel (d) represent parameter values used in panels (a)-(c).
   }   
\end{center}
\end{figure}
\clearpage

\subsection*{Supplementary Note: Comparison to other opinion dynamics model}

A large body of research has examined opinion dynamics \cite{castellano2009statistical}. Among the most widely used approaches are agent-based models, including the Bounded Confidence Models (BCMs)~\cite{deffuant2000mixing,axelrod2021preventing,sasahara2021social} and the Friedkin--Johnsen (FJ) model~\cite{friedkin1990social}. Although these models can reproduce polarized opinion states (PO states), the mechanisms underlying polarization differ from those investigated here. In BCMs, polarization emerges primarily through adjustments to the initial distribution of opinions and/or the network structure. In the FJ model, by contrast, polarization results from heterogeneity in individuals’ internal opinion parameters (i.e., beliefs). Importantly, in both models polarization tends to persist even when social interactions are weakened or links are removed; conversely, adding links has been shown to reduce polarization~\cite{zhu2021minimizing}.
In our model, however, the PO state is driven by the weights of the network, i.e., the strength of interactions among users. This implies that interventions targeting interaction strengths can effectively shift the system from a polarized state (PO) to a relatively aligned state (RA). Our findings therefore highlight the potential for social network interventions to exert substantial influence on polarization dynamics, even without relying on complex targeted interventions. A key direction for future work is to develop methods for inferring which model type—agent-based models or our proposed framework--best captures polarization patterns observed in empirical social network data.
\end{document}